\documentclass[journal]{IEEEtran}
\usepackage[T1]{fontenc}   
\usepackage{microtype}     
\usepackage{textcomp}      
\usepackage[dvipsnames]{xcolor}
\usepackage{cite}
\usepackage{amsmath, amssymb}
\allowdisplaybreaks
\usepackage{algorithmic}
\usepackage{array}
\usepackage{url}
\usepackage{subcaption}
\usepackage[colorlinks=true, linkcolor=blue, citecolor=blue, urlcolor=blue]{hyperref}
\usepackage{graphicx}
\usepackage[inkscapelatex=false]{svg}
\usepackage{titlesec}
\titlespacing*{\section}{0pt}{*0.8}{*0.6} 
\titlespacing*{\subsection}{0pt}{*0.6}{*0.4} 
\usepackage{orcidlink}
\usepackage[normalem]{ulem}

\IEEEoverridecommandlockouts

\makeatletter
\def\@IEEEpubidpullup{7.5\baselineskip}
\let\@IEEE@save@titlepagestyle\ps@IEEEtitlepagestyle
\def\ps@IEEEtitlepagestyle{%
  \@IEEE@save@titlepagestyle
  \def\@oddfoot{\hbox to\textwidth{\@IEEEfooterstyle\footnotesize\raisebox{\footskip}[0pt][0pt]{\parbox[b]{\columnwidth}{\raggedright\@IEEEpubid}}\hfill}}%
  \let\@evenfoot\@oddfoot
}
\makeatother
\IEEEpubid{\copyright 2026 IEEE. Personal use of this material is permitted. Permission from IEEE must be obtained for all other uses, in any current or future media, including reprinting/republishing this material for advertising or promotional purposes, creating new collective works, for resale or redistribution to servers or lists, or reuse of any copyrighted component of this work in other works.\newline DE~Zerrari, A.~Bendaimi, and H.~Arslan, ``Backward-Compatible Tag-Based PHY Authentication for Secure WLAN Sensing,'' \textit{IEEE Wireless Commun. Lett.}, vol.~15, pp.~3636--3640, 2026, doi: \url{https://doi.org/10.1109/LWC.2026.3697982}.}

\begin{document}

\title{Backward-Compatible Tag-Based PHY Authentication for Secure WLAN Sensing}

\author{Dhia~Elhak~Zerrari\,\orcidlink{0009-0006-9328-5560},
        Amira~Bendaimi\,\orcidlink{0009-0001-2708-5425},
        and~H\"{u}seyin Arslan\,\orcidlink{0000-0001-9474-7372},~\IEEEmembership{Fellow,~IEEE}
\thanks{\scriptsize Dhia Elhak Zerrari, Amira Bendaimi and Hüseyin Arslan are with the Department of Electrical and Electronics Engineering, Istanbul Medipol University, Istanbul, 34810, Turkey (email: dhia.zerrari@std.medipol.edu.tr, amira.bendaimi@std.medipol.edu.tr, huseyinarslan@medipol.edu.tr).}   
}
\maketitle

\begin{abstract}
Driven by ongoing standardization efforts, Wi-Fi sensing has recently emerged as a promising functionality within wireless local area networks (WLANs). However, due to the use of standard null-data packets (NDPs) to estimate the channel, conventional Wi-Fi sensing systems are susceptible to spoofing attacks. To counter this threat, we propose a lightweight tag-embedding scheme that enables physical layer authentication (PLA) of sensing signals. Specifically, a secret tag is superimposed onto the orthogonal frequency division multiplexing (OFDM) channel sounding signal, allowing legitimate WLAN receivers to verify signal authenticity and detect spoofed transmissions. A unique tag is generated for each transmission to avoid replay attacks, and the subcarrier occupancy of each tag is randomized to minimize the mean channel estimation error at each subcarrier. Simulation results demonstrate reliable authentication performance with a $95\%$ correct authentication rate at a signal-to-noise ratio (SNR) of $10\,\mathrm{dB}$, while introducing an additional channel estimation error of only $0.52\,\mathrm{dB}$ at legacy receivers that are unaware of the embedded tag. Thus, our proposed scheme provides an effective and backward-compatible countermeasure against spoofing attacks, enhancing the trustworthiness of future Wi-Fi sensing systems.
\end{abstract}
\IEEEpeerreviewmaketitle

\begin{figure*}[t!]
\centering
\begin{minipage}{.45\textwidth}
  \centering
    \includegraphics[width=0.85\linewidth]{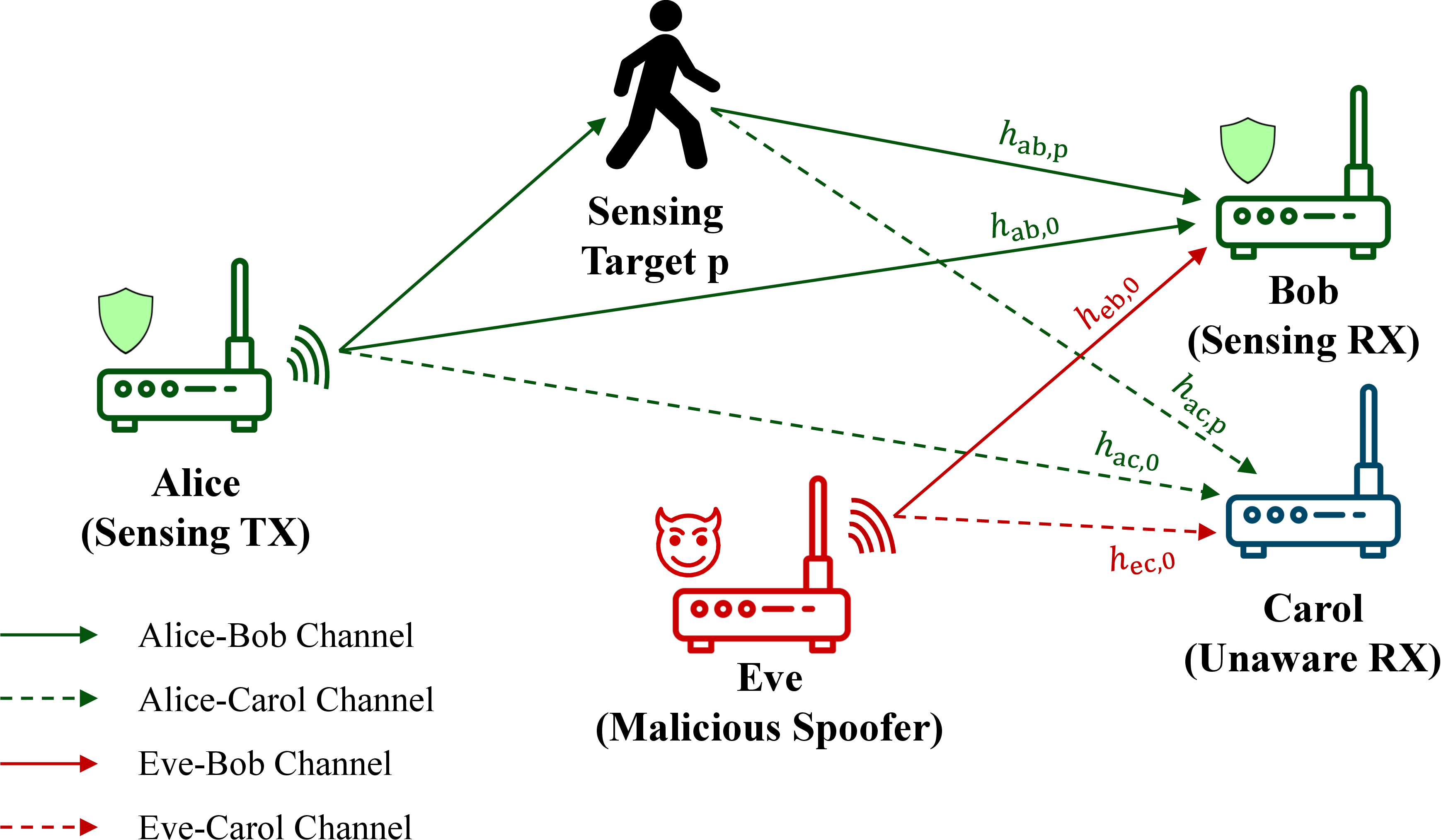}
    \caption{ISAC system model under a spoofing attack.}
    \vspace{-6pt}
    \label{fig:sys_model}
\end{minipage}\hfill%
\begin{minipage}{.45\textwidth}
    \centering
    \includegraphics[width=0.85\linewidth]{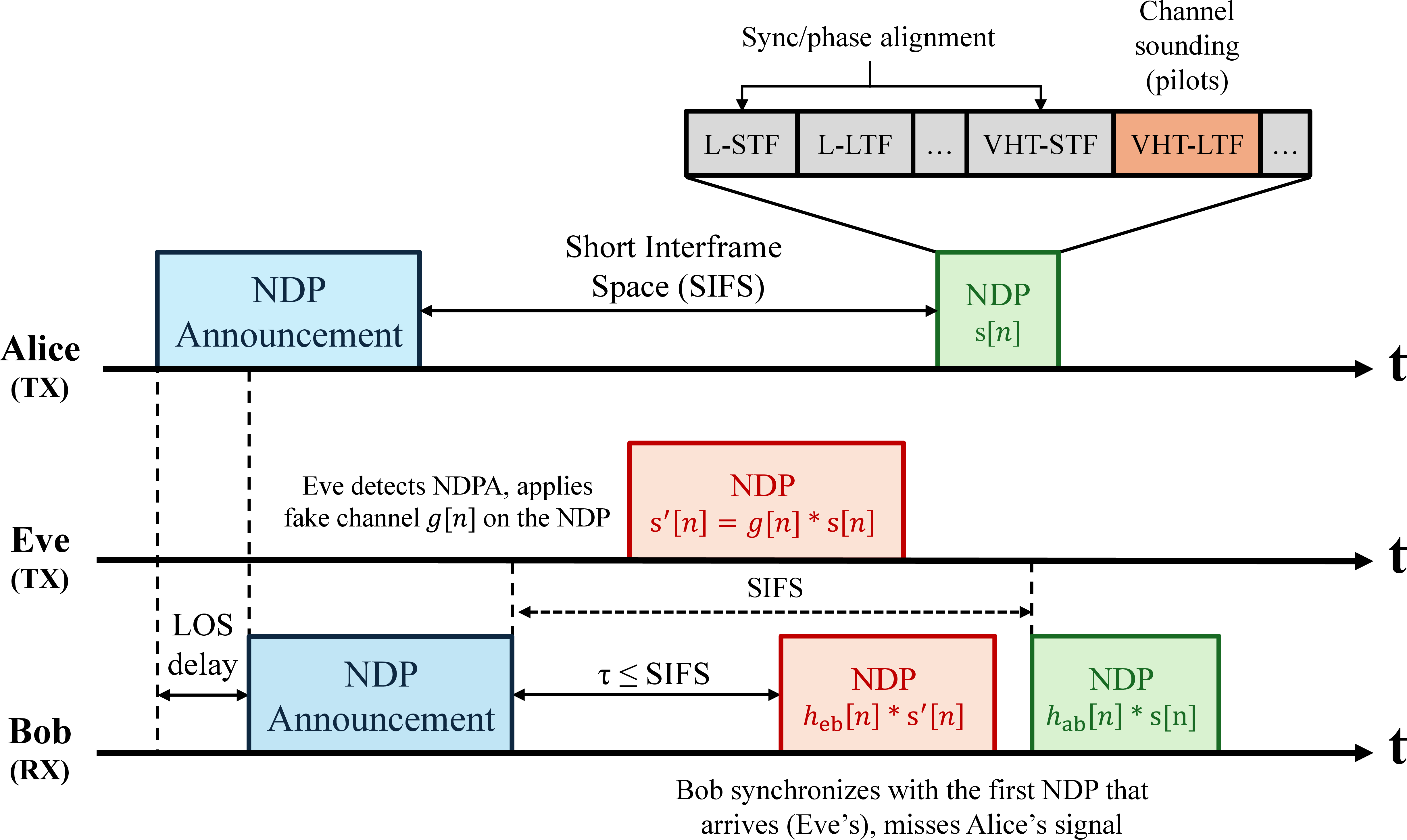}
    \caption{Spoofing attack strategy.}
    \vspace{-6pt}
    \label{fig:ndp_attack_timeline}
\end{minipage}%
\end{figure*}
\section{Introduction}
\IEEEPARstart{I}{ntegrated} sensing and communication (ISAC) is an emerging technology that allows the reuse of existing communication equipment, such as that used for cellular or Wi-Fi, for RF sensing purposes \cite{wifi-isac-survey-2022}. The technology possesses great potential in several fields including home automation and security \cite{pu-wisee-2013}, medical monitoring \cite{wang-wifall-2017} and human identification \cite{wang-huID-2022}. These applications have attracted the interest of the WLAN standards committee, which has recently published the 802.11bf amendment that aims to unify WLAN-based sensing procedures \cite{du-xu-802.11bf-overview-2024}. The 3rd Generation Partnership Project (3GPP) is also considering ISAC as a key paradigm for the design of future mobile networks, starting with the upcoming 6G standard \cite{aldirmaz-6g-ISAC}.

Although ISAC-based sensing shows great promise, industry leaders and consumers are concerned about the potential for privacy invasion and threats of both electronic (RF) and cyber attacks \cite{martins2025delving}. Unlike traditional radar systems, ISAC devices are still in their commercial infancy and have yet to be heavily scrutinized from a security standpoint. Current implementations of ISAC using Wi-Fi are particularly vulnerable to spoofing due to the reuse of standard, publicly known training sequences for channel sensing \cite{yildirim2025ofdm}. Spoofing attacks deceive sensing devices into detecting channel signatures that are not truly present in the environment. For instance, an attacker could conceal anomalous targets in home security systems, enabling intrusions. RF-based human authentication systems could also be compromised by replaying recorded signal reflections \cite{2024-chrysanidis-replay}.

Several strategies have been proposed to mitigate spoofing attacks. For instance, the authors in \cite{ali2025cooperative} utilized cooperative multi-nodal sensing networks to cross-validate sensing results and detect any disparity, at the expense of adding significant complexity and cost. In contrast, the authentication schemes outlined in \cite{sankhe-rfimpair-2020, tian2026-cfo-pla-isac} rely solely on existing RF hardware impairments as a unique fingerprint to differentiate legitimate transmitters from attackers without requiring redundancy. However, the difficulty of extracting consistent hardware features can lead to unsatisfactory authentication rates. Furthermore, in \cite{tian2026-cfo-pla-isac, liu-csi-pla-2018} the channel state information (CSI) itself is used for authentication by detecting abnormal drifts in the channel's impulse response (CIR) or its frequency response (CFR) over time. Nonetheless, this is unsuited for environments with time-varying channels, as is the case in practical sensing scenarios. Finally, in \cite{janjua-pilot-alloc-2022}, the placement of pilots in the sounding signal is randomized in a manner only known to the legitimate devices. Yet, this approach does not address the issue of compatibility with legacy devices. 

In light of the aforementioned shortcomings, this letter proposes a novel backward-compatible tag-based PLA mechanism specifically designed for WLAN-based ISAC systems under spoofing attacks. The scheme is based on embedding secret tag signals onto the OFDM channel sensing signals, enabling legitimate WLAN receivers to verify signal authenticity. Moreover, to prevent replay attacks, a unique tag is generated per packet, and its subcarrier allocation is randomized on each transmission, which leads to the distribution of estimation error across the spectrum while reducing the likelihood of tag recovery by a malicious adversary. 
To analytically assess the robustness of the proposed scheme, closed-form expressions for the false alarm and detection probabilities are derived. Finally, simulation results confirm reliable detection with a correct authentication rate of $95\%$ at $10\,\mathrm{dB}$ SNR, while legacy receivers experience only a marginal increase in channel estimation error. Consequently, the presented approach offers robust spoofing detection without disrupting interoperability with legacy WLAN devices. 
\section{System Model and Problem Formulation} \label{prob-statement}
Consider an OFDM-based bi-static ISAC system as depicted in Fig.~\ref{fig:sys_model}. It consists of an authenticated WLAN transmitter and receiver, denoted as Alice and Bob, respectively, along with an adversary Eve, who attempts to launch a spoofing attack. Eve aims to compromise the sensing integrity by deceiving Bob into detecting fake targets at false range-velocity coordinates while potentially concealing the true targets. During each sensing instance, Alice transmits a standard NDP signal, denoted by $s[n]$, which Bob uses to estimate the channel. According to the WLAN standard, however, Alice must first send an NDP Announcement (NDPA) frame before the actual NDP transmission. As highlighted in~\cite{yildirim2025ofdm}, this mandatory announcement exposes the sensing process to spoofing attacks. Upon detecting the NDPA from Alice, Eve opportunistically transmits a forged NDP signal $s'[n] = g[n] * s[n]$ as illustrated in Fig.~\ref{fig:ndp_attack_timeline}, where $g[n]$ represents an artificial channel intentionally designed to mislead Bob's estimation.
 
 
To capture the propagation characteristics under legitimate and spoofed scenarios, the corresponding time-frequency channel model is expressed as
\begin{equation}
    H_{xy}[k,m] = \sum_{p=0}^{P_{xy}-1} \alpha_p^{xy} e^{-j2\pi k \frac{\tau_p^{xy}}{N T_s}} e^{j2\pi f_p^{xy} m T_\text{PRI}},
\end{equation} 
where $x \in \{a,e\}$ and $y \in \{b,c\}$ denote the transmitter (Alice or Eve) and receiver (Bob or Carol), respectively. The model sums over $P_{xy}$ propagation paths, each with complex gain $\alpha_p^{xy}$, delay $\tau_p^{xy}$, and Doppler shift $f_p^{xy}$, where $p=0$ corresponds to the line-of-sight (LOS) component. Here, $k$ and $m$ are the OFDM subcarrier and NDP pulse indices, $N$ is the number of subcarriers, $T_s$ the sampling interval, and $T_\text{PRI}$ the pulse repetition interval. Doppler effects are assumed constant over one NDP packet. The purpose of the proposed PLA scheme is to protect the sensing process at Alice and Bob from Eve's spoofing attacks. Additionally, as shown in Fig.~\ref{fig:sys_model}, an unaware legacy receiver (Carol) is considered to evaluate the compatibility of the scheme with a legacy WLAN device. 
\section{Proposed WLAN-TBPLA Framework} \label{proposed-tag}
\vspace{-2pt}
\begin{figure*}[ht!]
    \centering
    \includegraphics[width=0.9\linewidth]{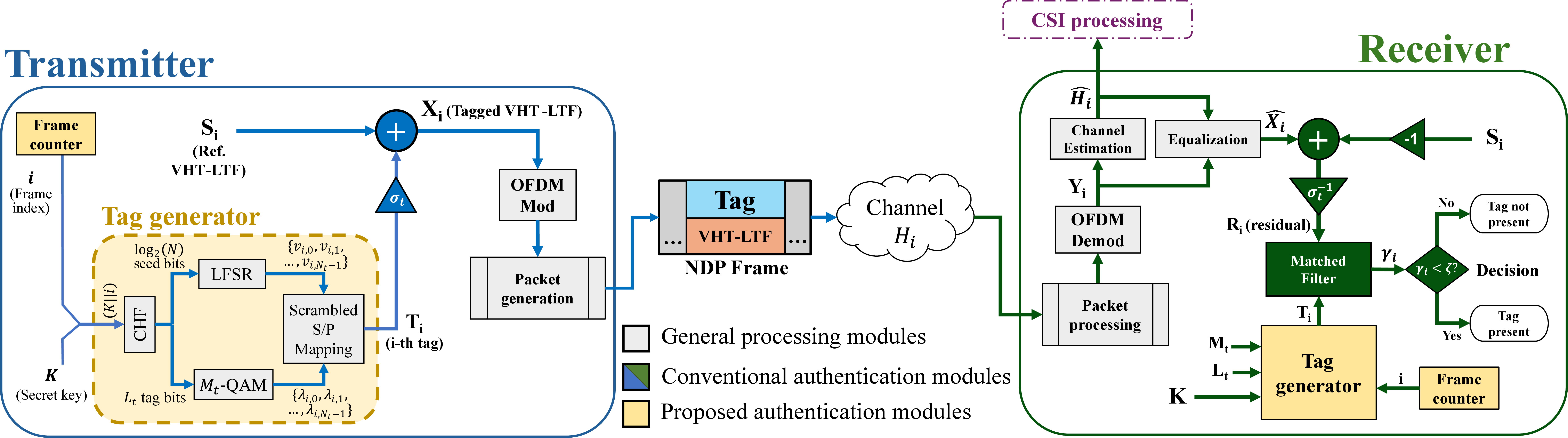}
    \caption{Block diagram of the proposed WLAN-TBPLA framework.}
    \vspace{-14pt}
    \label{fig:tx_rx_blocks}
\end{figure*}
This section presents the proposed WLAN-TBPLA framework, depicted in Fig.~\ref{fig:tx_rx_blocks}. It operates in two main stages: tag generation and embedding at the transmitter, followed by tag extraction and verification at the receiver. In each sensing frame, Alice places a tag on its NDP packets such that Bob is able to discern their legitimacy, thereby differentiating them from those forged by Eve. Initially, Alice and Bob agree on a shared secret key $K$, which is generated using physical layer features \cite{aldaghri-chkey-2020}, or negotiated through higher-level key exchange protocols \cite{ieee80211be2024} \footnote{$K$ can be established using existing WLAN security mechanisms (WPA2/3) \cite{ieee80211be2024}. The proposed scheme is agnostic to the key establishment method, as it does not affect the rest of the frame authentication operation.}.  A common set of tag-embedding parameters is also established, namely the tag modulation order, $M_t$, the tag scaling factor, $\sigma_t$, and the tag length in bits, $L_t$. This tag authentication process is performed in every frame throughout the sensing session. Notably, Eve is capable of intercepting all NDPAs and NDPs at high SNR, and has full knowledge of the PLA scheme except for $K$. Due to spatial de-correlation, Eve is unable to obtain the legitimate channel response $H_{ab}$.
\subsection{Tag Generation at the Transmitter}
Consistent with standard tag authentication schemes \cite{xie2020survey}, the tag generation process at the transmitter begins by producing a tag bit-stream derived from the shared secret key $K$. However, due to the lack of a message bit-stream, a unique frame marker $i$ is appended to $K$ to ensure that each tagged frame yields a distinct bit-stream to prevent replay attacks. Thus, the tag bit-stream of the $i$-th frame is given by $t_i = \text{CHF}(K || i)$, where $\text{CHF}$ is a cryptographic hash function and $\cdot\,||\,\cdot$ denotes the bit concatenation operator. The frame identifier $i$ is a shared synchronized counter that can be derived from the control fields introduced in the 802.11bf WLAN standard, such as the measurement session and exchange IDs \cite{du-xu-802.11bf-overview-2024}. The use of the frame identifier $i$ to generate bit-streams prevents replay attacks even if it is intercepted by an eavesdropper, as long as each value of $K||i$ is only used once. 

The first $L_t$ bits of $t_i$ are then modulated using an $M_t$-ary modulation, resulting in a set of tag symbols $\Lambda_i = \{\lambda_{i,0}, \lambda_{i,1},...,\lambda_{i,N_t-1}\}$, where $N_t = \lceil\frac{L_t}{\log_2{M_t}}\rceil$. The next $\log_2(N)$ bits of $t_i$ are used as a seed $v_{i,0}$ for a maximum-length linear feedback shift register (LFSR), and the next $N_t$ values of the LFSR sequence form a pilot index sequence $V_i = \{v_{i,0}, v_{i,1}, ...,v_{i,N_t-1}\}$. Moreover, the tag signal $T_i$ is composed in the frequency domain as follows
\begin{equation}
    T_i[k] = \begin{cases}
                \Lambda_i(\kappa), & k = V_i(\kappa), \\
                0, & \text{otherwise},
             \end{cases}
\end{equation}
where $\kappa \in \{0, 1, ..., N_t -1\}$. The tagged channel training signal $X_i$ is constructed as
\begin{equation}
\label{eqn:x-def}
    X_i[k] = S_i[k] + \sigma_t T_i[k],
\end{equation}
where $S_i$ is the reference NDP signal in the frequency domain. For compatibility, the tag is embedded exclusively on the Very High Throughput Long Training Field (VHT-LTF) section of the NDP. \footnote{Although this study considers the IEEE 802.11ac VHT preamble, the proposed scheme can be readily extended to newer and equally vulnerable variants, such as High Efficiency (HE, 802.11ax) and Extremely High Throughput (EHT, 802.11be \cite{ieee80211be2024}).} Consequently, $S_i$ is assumed to consist of a single OFDM symbol representing the VHT-LTF. The tagged signal $X_i$ is transmitted by Alice instead of $S_i$, thus allowing Bob to verify the authenticity of the received NDP.

Since $v_{i,0}$ is computationally indistinguishable from a uniform draw over $N_s$ active subcarriers, Eve's single-frame probability of guessing $V_i$ is $\Pr\{\hat{V}_i = V_i\} \leq \frac{1}{N_s}$. Pseudorandomness ensures past observations yield no predictive advantage for $V_{i+1}$, neutralizing accumulation-based attacks.

\subsection{Authentication at the Receiver}
At the receiver side, authentication is performed by estimating the channel from the received tagged NDP and determining whether a valid tag is present. Assuming perfect time and frequency synchronization, the received channel training signal $Y_i$ is given in the frequency domain by
\begin{equation}
    Y_i[k] = H_i[k] \odot X_i[k] + W_i[k],
\end{equation}
where $\odot$ represents the Hadamard product and $W_i$ is additive white Gaussian noise (AWGN) with distribution $\mathcal{CN}(0, \sigma_w^2 I_{N})$. The channel coefficients are first calculated at the untagged subcarriers, and then interpolated across the remaining indices, resulting in a rough channel estimate $\hat{H}$ as
\begin{equation}
\label{eqn:H-hat-def}
    \hat{H}_i[k] =
    \begin{cases}
        \dfrac{Y_i[k]}{S_i[k]}, & k \notin V_i, \\[6pt]
        \mathcal{I}\big\{ \hat{H}_i[m]:_{m \notin V_i} \big\}(k), & \text{otherwise,}
    \end{cases}
\end{equation}
where $\mathcal{I}\{\cdot\}$ is the interpolation/extrapolation operator. The signal estimate $\hat{X}_i$ is obtained through the equalization of $Y_i$ as follows
\begin{equation}
    \begin{split}
    \hat{X}_i[k] &= Y_i[k] \oslash \hat{H}_i[k]\\
                &=   X_i[k] (1 - \Xi_i[k] \oslash \hat{H}_i[k]) + W_i[k] \oslash  \hat{H}_i[k]\\
                & \approx X_i[k] + W_i'[k],
    \end{split}
\end{equation}
where $\oslash$ denotes the Hadamard division, $\Xi_i[k] = \hat{H}_i[k] - H_i[k]$ is the channel estimation error and for each subcarrier $k$, $W_i'[k] \sim \mathcal{CN}(0, \frac{\sigma_w^2}{|\hat{H}_i[k]|^2})$. A residual $R_i$ is then distilled from $\hat{X}_i$ as follows
\begin{equation}
    R_i[k] = \frac{1}{\sigma_t} \left(\hat{X}_i[k] - S_i[k]\right).
\end{equation}
Based on $R_i$, a binary hypothesis test statistic $\gamma_i$ is evaluated to classify the signal into one of the two possibilities
\[
\begin{aligned}
\mathcal{H}_0 &: \text{Valid tag is not present in the residual,} \\
\mathcal{H}_1 &: \text{Valid tag is present in the residual.}
\end{aligned}
\]
For the receiver to discern the authenticity of the tag, it must possess its own copy of the tag signal $T_i$. The statistic $\gamma_i$ is subsequently obtained by applying a matched filter as follows
\begin{equation}
    \label{eqn:gamma-def}
    \begin{split}
        \gamma_i & =  \Re \{ T_i^H R_i \} = \frac{1}{\sigma_t}\Re \left\{ T_i^H  \left(\hat{X}_i - S_i\right)\right\}\\
             & = \begin{cases}
                \mathcal{H}_0: \frac{1}{\sigma_t} \Re\{T_i^H W_i'\},\\
                \mathcal{H}_1: \|T_i\|^2 + \frac{1}{ \sigma_t} \Re\{T_i^H W_i'\}.
             \end{cases} \\
    \end{split}
\end{equation}
Finally, Bob makes a decision on the authenticity of the signal by comparing $\gamma_i$ to a threshold $\zeta$
\begin{equation}
    \begin{split}
        \gamma_i < \zeta \longrightarrow \textit{Decide }\mathcal{H}_0 \\
        \gamma_i \geq \zeta \longrightarrow \textit{Decide }\mathcal{H}_1
    \end{split}
\end{equation}
If the received signal is authentic, Bob improves its initial channel estimate by treating the tagged symbols as additional pilots, and evaluating the channel response at their subcarriers. The threshold $\zeta$ is chosen based on the desired authentication performance with respect to the system parameters.
\section{Performance Analysis of the Proposed Scheme} \label{perf-analysis}
To assess the robustness of the proposed scheme, the probabilities of false alarm and correct detection are expressed respectively as $P_{FA} = \Pr\{\gamma_i \ge \zeta \mid \mathcal{H}_0\}$ and $P_{auth} = \Pr\{\gamma_i \ge \zeta \mid \mathcal{H}_1\}$. The corresponding probability density functions (PDFs) can be integrated as
\begin{equation}\small
\label{eqn:pfa-pd-def}
  P_{FA} = \int_{\zeta}^{+\infty} p_{\gamma_i \mid \mathcal{H}_0}(x)\,dx, \qquad
  P_{auth}  = \int_{\zeta}^{+\infty} p_{\gamma_i \mid \mathcal{H}_1}(x)\,dx.
\end{equation}
From \eqref{eqn:gamma-def}, it is evident that $\gamma_i \mid \mathcal{H}_0$ and $\gamma_i \mid \mathcal{H}_1$ each obey Gaussian distributions, denoted $\mathcal{N}(\mu_0, \sigma_0^2)$ and $\mathcal{N}(\mu_1, \sigma_1^2)$ respectively; thus the integral for $P_{FA}$ yields a Q-function, $Q\!\left((\zeta-\mu_0) / \sigma_0\right)$, where
\begin{equation}
\begin{split}
    \mu_0 &= E\!\left[\frac{1}{\sigma_t} \Re\{T_i^H W_i'\}\right] = 0,
\end{split}
\end{equation}
and
\begin{equation}
    \begin{split}
        \sigma_0 &= \sqrt{E\!\left[\left|\frac{1}{\sigma_t}\Re\{T_i^H W_i'\}\right|^2\right]} \\
        &= \sqrt{ \frac{1}{2\sigma_t^2} T_i^H \mathrm{diag}\!\left(\frac{\sigma_w^2}{|\hat{H}_{eb}[k]|^2}\right) T_i } \\
        &= \frac{\sigma_w}{\sigma_t\sqrt{2}}\sqrt{ \sum_{k=0}^{N-1} \frac{|T_i[k]|^2}{|\hat{H}_{eb}[k]|^2}}.
    \end{split}
\end{equation}
To enforce a maximum false alarm rate $\epsilon_{FA}$, the corresponding detection threshold is found through the inverse Q-function as $\zeta_{FA} = \sigma_0 Q^{-1}(\epsilon_{FA})$.
From \eqref{eqn:pfa-pd-def}, evaluating the integral for $P_{auth}$ results in $Q\!\left((\zeta - \mu_1) / \sigma_1 \right)$, where

\begin{equation}
    \begin{split}
        \mu_1 &= E\!\left[\|T_i\|^2+\frac{1}{\sigma_t} \Re\{T_i^H W_i'\}\right] = \|T_i\|^2,
    \end{split}
\end{equation}
and

\begin{equation}
    \label{eq:sigma1}
    \begin{split}
        \sigma_1 &= \sqrt{\text{Var}\left( \|T_i\|^2 + \frac{1}{\sigma_t} \Re\{T_i^H W_i'\} \right)} \\
        &= \sqrt{\frac{1}{2\sigma_t^2} T_i^H \mathrm{diag}\left(\frac{\sigma_w^2}{|\hat{H}_{ab}[k]|^2}\right) T_i} \\
        &= \frac{\sigma_w}{\sigma_t\sqrt{2}} \sqrt{ \sum_{k=0}^{N-1} \frac{|T_i[k]|^2}{|\hat{H}_{ab}[k]|^2}}
    \end{split}
\end{equation}
Substituting the threshold $\zeta = \zeta_{FA}$ determined from the false alarm constraint gives $P_{auth,\epsilon} = Q\!\left( (\zeta_{FA} - \|T_i\|^2) / \sigma_1 \right)$.


For the sake of comparison with the existing tag-based PLA \cite{yu2008-sup-pla}, $\sigma_t^2$ can be redefined such that a constant power allocation is maintained for the tag signal relative to the reference signal, by imposing the constraint { $\rho_s^2 + \rho_t^2 = 1$}, thus
\begin{subequations}
    \begin{equation}
    \label{eqn:s-pow}
        \|S_i\|^2 = \rho_s^2 \|X_i\|^2,
    \end{equation}
    \begin{equation}
    \label{eqn:t-pow}
        \sigma_t^2\|T_i\|^2 = \rho_t^2 \|X_i\|^2,
    \end{equation}
\end{subequations}
where $\rho_s^2$, $\rho_t^2$ are the power allocation factors of the reference signal and the tag signal, respectively. It is assumed that $E[S_i^H T_i] = 0$. Substituting \eqref{eqn:s-pow} into \eqref{eqn:t-pow} yields
\begin{equation}
\label{eqn:sigma-rho-def}
    \begin{split}
        \sigma_t^2 & =  \frac{\rho_t^2}{1-\rho_t^2} \frac{\|S_i\|^2}{\|T_i\|^2}. \\
    \end{split}
\end{equation}
When $\sigma_t$ is varied independently, it directly controls the symbol spacing within the tag signal $T_i$. However, under the constraint in \eqref{eqn:sigma-rho-def}, $\sigma_t$ becomes implicitly governed by the total energy of $T_i$. Moreover, if the tag symbols $\Lambda_i$ are generated with a constant power envelope, the symbol spacing becomes inversely proportional to the total number of symbols $N_t$.

To assess backward compatibility, the normalized mean square error (NMSE) of Carol's channel estimate is derived as follows (assuming $\|T_i\|^2 = N_t$, $\|S_i\|^2 = N_s$, $E[|H_{ac}|^2] = 1$)

\begin{equation}\footnotesize
    \begin{split}
        \text{NMSE} &= E\left[ |\hat{H}_{ac}[k]- H_{ac}[k]|^2 \right] = E\left[\bigg| \frac{Y_i[k]}{S_i[k]} - H_{ac}[k]\bigg|^2\right]\\
        &= E\left[\bigg| \frac{H_{ac}[k] (S_i[k] + \sigma_t T_i[k])}{S_i[k]} + \frac{W_i[k]}{S_i[k]} - H_{ac}[k]\bigg|^2\right], \\
        &= E\left[ \bigg| \frac{\sigma_t H_{ac}[k] T_i[k]}{S_i[k]} + \frac{W_i[k]}{S_i[k]} \bigg|^2 \right] = \sigma_t^2 \frac{N_t}{N_s} + \sigma_w^2.
    \end{split}
\end{equation}

We define the maximum tolerable NMSE floor $\Delta_{max}$ (in the limit $\sigma_w^2 \rightarrow 0$) as $\Delta_{max} = \sigma_t^2 N_t / N_s = \sigma_t^2 L_t / (\log_2(M_t) N_s)$.
This establishes a design rule where $\sigma_t^2$ and $L_t$ are chosen such that $\sigma_t^2 L_t / (\log_2(M_t) N_s) \leq \Delta_{max}$, ensuring a guaranteed performance floor for legacy devices.

Additionally, the proposed scheme incurs a minimal per-frame overhead of $\mathcal{O}(N_t)$ operations ($3N_t$ additions, $4N_t$ multiplications, and $N_t$ interpolations). This is significantly lighter than machine-learning-based passive PLAs \cite{sankhe-rfimpair-2020}. Furthermore, per-frame authentication eliminates block-processing delays, ensuring low latency for resource-constrained devices.

\section{Simulation Results}
\begin{figure*}[t!]
\centering
\begin{minipage}{0.48\textwidth}
  \centering
  \includegraphics[width=0.85\linewidth]{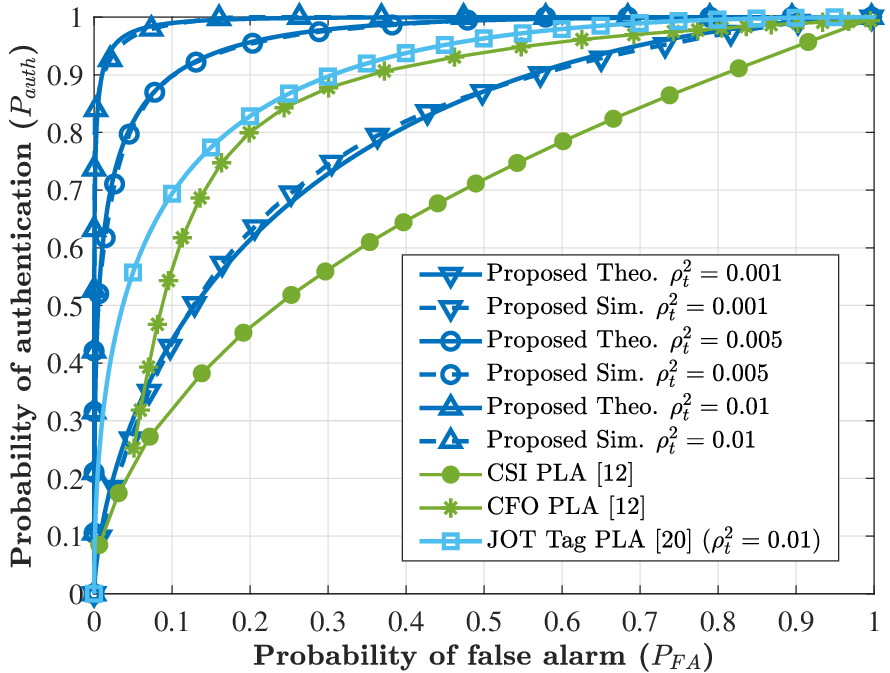}
  \vspace{-4pt}
  \caption{ROC comparison with related works.}
  \label{fig:roc_tag_pow}
\end{minipage}\hfill
\begin{minipage}{0.48\textwidth}
  \centering
  \includegraphics[width=0.85\linewidth]{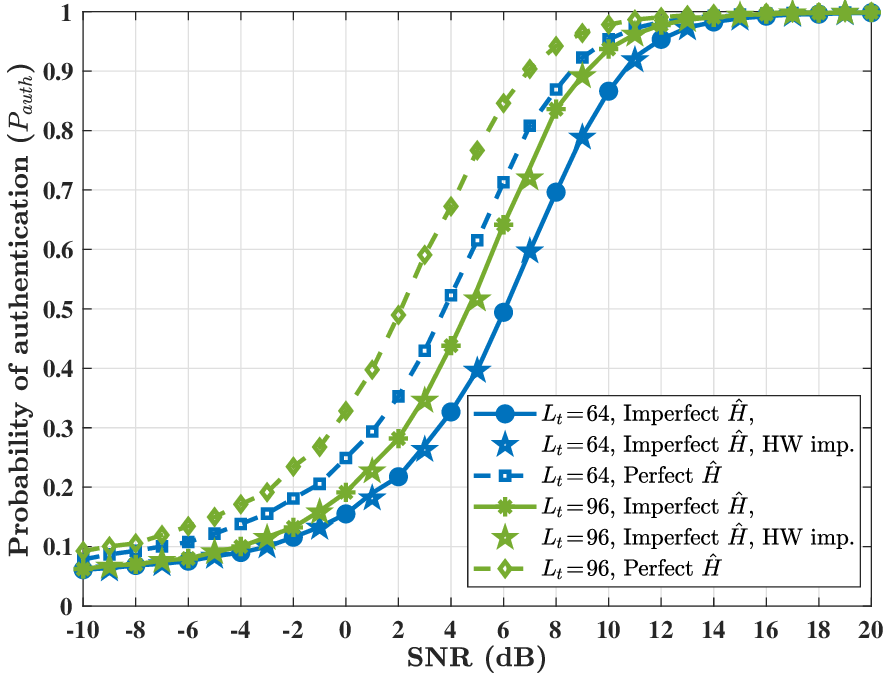}
  \vspace{-4pt}
  \caption{Authentication performance analysis.}
  \label{fig:pd_snr_tag_L}
\end{minipage}

\vspace{1em} 

\begin{minipage}{0.48\textwidth}
  \centering
  \includegraphics[width=0.85\linewidth]{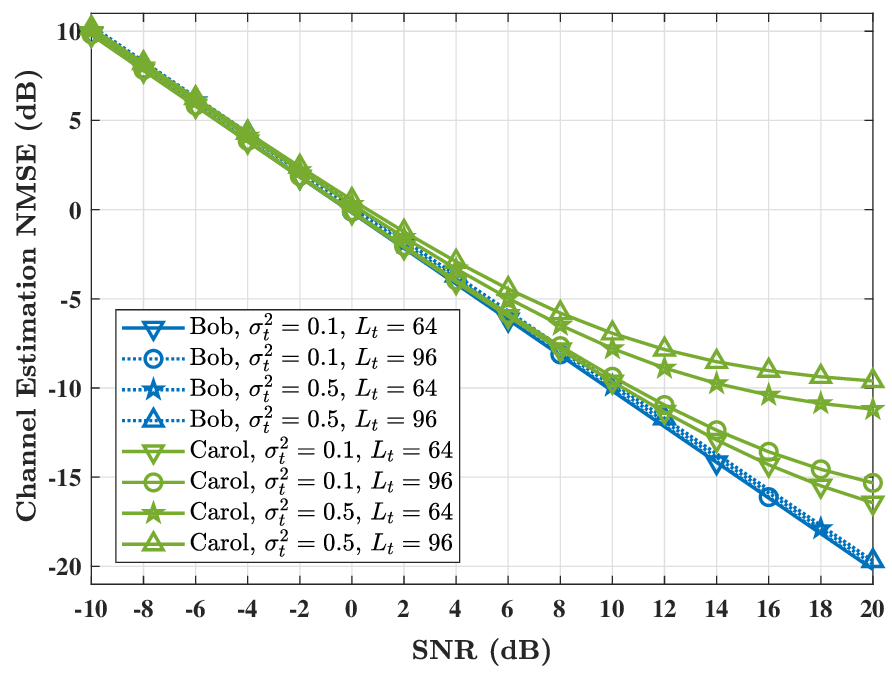}
  \vspace{-4pt}
  \caption{Channel estimation NMSE performance analysis.}
  \label{fig:bob_carol_chan_nmse}
\end{minipage}\hfill
\begin{minipage}{0.48\textwidth}
  \centering
  \includegraphics[width=0.85\linewidth]{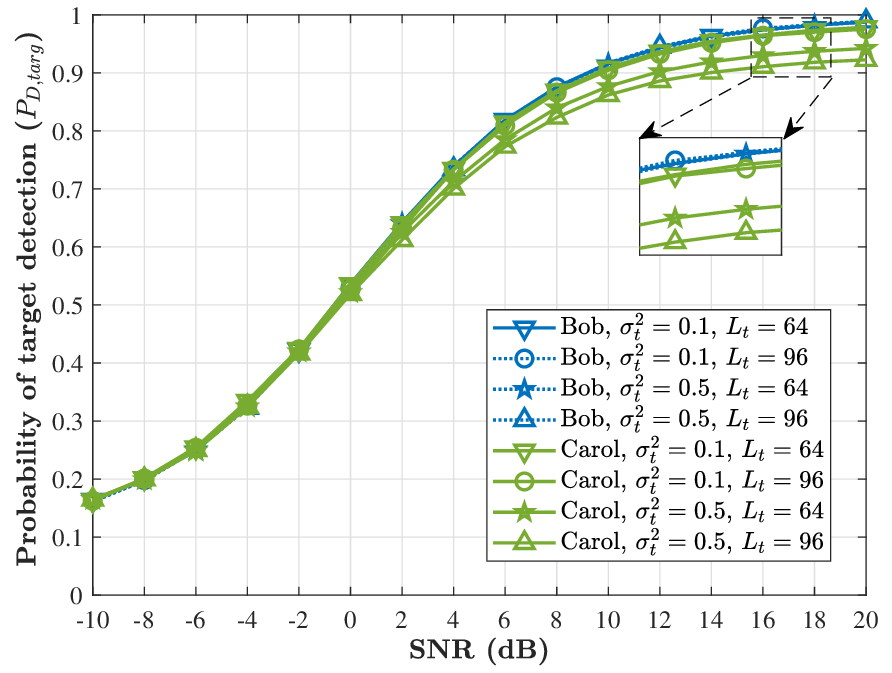}
  \vspace{-4pt}
  \caption{CFAR detection performance analysis.}
  \label{fig:bob_carol_cfar_rdm}
\end{minipage}

\vspace{-8pt}
\end{figure*}

To evaluate the authentication scheme, a 5~GHz IEEE 802.11ac WLAN system has been configured in 80~MHz mode spread over $N=256$ OFDM subcarriers, of which $N_s = 242$ subcarriers are actively used for pilot transmission. The secret key $K$ is defined as 128 bits, zero-padded to align with the SHA-256 cryptographic hash function used to generate the tag bit-streams, which are subsequently modulated to 4-QAM symbols ($M_t = 4$). The channel frequency response is modeled according to the 14-tap standard WLAN Channel Model C \cite{ieee80211be2024}, with additional varying mobile sensing targets to induce Doppler. For performance evaluation, each tagging parameter is varied independently while the remaining parameters are maintained at their reference settings: $L_t = 64$, $\sigma_t^2 = 0.1$, $\epsilon_{FA} = 0.05$, and $\mathrm{SNR} = 10\,\mathrm{dB}$.

First, the authentication performance of the proposed scheme is assessed through the receiver operating characteristic (ROC) curve illustrated in Fig.~\ref{fig:roc_tag_pow}. The ROC shows favorable authentication performance that improves with the increase in $\rho_t^2$, consistent with the behavior observed in conventional tag-based PLA frameworks \cite{yu2008-sup-pla, tan2025asynchronous, xie2025-rcc-tag}. For instance, a tag power allocation of $\rho_t^2 =0.01$ is enough to guarantee an authentication accuracy of over $96\%$ at only $5\%$ false alarm rate. Notably, unlike the traditional tag-based PLA schemes designed for communication systems, the proposed approach is not affected by message-tag interference. Additionally, theoretical findings closely align with simulation results, demonstrating the accuracy of our theoretical model in characterizing authentication performance for $P_{FA}$ and $P_{auth}$. Furthermore, comparative analysis demonstrates that the proposed method significantly outperforms existing CFO-based, CSI-based \cite{tian2026-cfo-pla-isac}, and tag-based \cite{xie2025-rcc-tag} approaches. Specifically, given $\rho_t^2=0.01$ and $P_{FA}=5\%$ as a common operating point for all schemes in Fig.~\ref{fig:roc_tag_pow}, our scheme achieves $P_{auth}$ of over $96\%$, while existing tag-based methods achieve $P_{auth}$ of $55.7\%$ and less than $26\%$ for passive schemes. The performance gap is attributed to the use of active tagging and strategic placement of tag symbols on subcarriers with known pilot symbols, which mitigates the estimation and equalization errors inherent in methods that tag unknown data symbols.

The influence of the tag parameters on authentication performance is further investigated in Fig.~\ref{fig:pd_snr_tag_L}, which illustrates the authentication probability $P_{auth}$ versus SNR for different tag lengths $L_t$. Notably, $P_{auth}$ improves with increases in either the SNR or the tag length. 
Specifically, an improvement of approximately $7.1\%$ in $P_{auth}$ is observed when $L_t$ increases from $64$ to $96$ at $\mathrm{SNR}=10\,\mathrm{dB}$. Although a larger number of tag symbols makes the tag signal more distinguishable from noise, it also yields diminishing returns in authentication performance due to fewer untagged pilots, which can exacerbate channel estimation errors at low SNR. For instance, the performance gap between imperfect and perfect channel estimates increases from $21.84\%$ at $L_t=64$ to $23.35\%$ at $L_t=96$ for $\mathrm{SNR}=5\,\mathrm{dB}$. To evaluate hardware impairment resilience, we injected normalized residual time and frequency offset errors of $\delta=4$ and $\eta=0.02$, respectively. We observe negligible impact on authentication performance within this typical synchronization error range. This robustness stems from the pilots' ability to track impairments, enabling the equalizer to mitigate synchronization errors effectively.

Moreover, the compatibility of our proposed scheme with existing devices is evaluated by measuring the channel estimation performance, as well as range-Doppler map (RDM)-based moving target detection at both Bob and Carol. The two receivers are assumed to be co-located and hence observe the same channel, i.e., $H_{ac} = H_{ab}$.  Fig.~\ref{fig:bob_carol_chan_nmse} illustrates the NMSE of the channel estimates at both Bob and Carol with respect to SNR for different tagging parameters. At Bob, estimation accuracy improves with increasing SNR. However, the improvements are essentially independent of the tag parameters. At Carol, the estimation accuracy degrades as either $\sigma_t^2$ or $L_t$ increases, which is expected as both parameters affect the deviation of the tagged NDP signal from the standard NDP signal that Carol anticipates. This deviation is clearly reflected in Fig.~\ref{fig:bob_carol_cfar_rdm}, which demonstrates target detection accuracy $P_{D,targ}$ using an ordered statistic constant false alarm rate (OS-CFAR) detector for different SNR levels. Bob's detection performance remains invariant to the tag parameters, achieving a $P_{D,targ}$ of $91.5\%$ at an SNR of $10\,\mathrm{dB}$, while Carol experiences a measurable performance degradation relative to Bob, which becomes more pronounced at higher SNR levels. For instance, setting $L_t=96$ and $\sigma_t^2=0.5$ results in a detection accuracy loss floor of $6.62\%$ at Carol. Nevertheless, the additional estimation and target detection errors introduced by the tagging process remain small, and authenticated transmitters may further minimize them through conservative parameter selection when interoperability with legacy devices is desired. For instance, setting $L_t =64$ and $\sigma_t^2=0.1$ is sufficient to obtain a $P_{auth}$ exceeding $86.6\%$ at Bob, while only incurring an additional $0.52\,\mathrm{dB}$ of channel estimation error at Carol, as well as reducing the $P_{D,targ}$ floor gap to only $1.31\%$. Furthermore, parameter tuning may also enhance tag covertness; for instance, $\sigma_t$ could be increased in low SNR conditions to improve $P_{auth}$, and reduced under high-SNR conditions to remain stealthy. This dynamic adaptation can be readily integrated into WLAN systems by mapping tag parameters to the modulation and coding scheme (MCS) table as defined in the IEEE 802.11be standard \cite{ieee80211be2024}.
\section{Conclusion} \label{conclusion}
This letter proposes a novel tag-based physical layer authentication framework to secure WLAN sensing against spoofing attacks. By embedding secret tags into NDPs with subcarrier scrambling and adaptive power control, the proposed scheme achieves robust fake sensing signal detection while maintaining full backward compatibility. Owing to its lightweight design, the proposed scheme can be readily integrated into existing IEEE 802.11bf sensing systems without requiring significant protocol modifications. Future work will focus on experimental validation on an SDR-based IEEE 802.11ac/bf testbed, extending the threat model to realistic attack conditions including signal superposition and near-far power disparities.
 \section*{Acknowledgment}
This work was supported by the Scientific and Technological Research Council of Turkey (TÜBİTAK) under Grant Number 123E226.
\ifCLASSOPTIONcaptionsoff
  \newpage
\fi


\bibliographystyle{IEEEtran}
\bibliography{references}

@ARTICLE{wifi-isac-survey-2022,
  author={Tan, Sheng and others},
  journal={IEEE Internet Things J.}, 
  title={Commodity {WiFi} Sensing in Ten Years: Status, Challenges, and Opportunities}, 
  year={2022},
  volume={9},
  number={18},
  pages={17832-17843},
  doi={10.1109/JIOT.2022.3164569}}

@inproceedings{pu-wisee-2013,
author = {Pu, Qifan and others},
title = {Whole-home gesture recognition using wireless signals},
year = {2013},
isbn = {9781450319997},
publisher = {Association for Computing Machinery},
address = {New York, NY, USA},
doi = {10.1145/2500423.2500436},
booktitle = {Proc. of the 19th Annual Intl. Conference on Mobile Computing \& Networking},
pages = {27–38},
numpages = {12},
location = {Miami, Florida, USA},
series = {MobiCom '13}
}

@ARTICLE{wang-wifall-2017,
  author={Wang, Yuxi and others},
  journal={IEEE Trans. Mobile Comput.}, 
  title={{WiFall}: Device-Free Fall Detection by Wireless Networks}, 
  year={2017},
  volume={16},
  number={2},
  pages={581-594},
  doi={10.1109/TMC.2016.2557792}}

@ARTICLE{wang-huID-2022,
  author={Wang, Dazhuo and others},
  journal={IEEE Internet Things J.}, 
  title={{CAUTION: A Robust WiFi-Based Human Authentication System via Few-Shot Open-Set Recognition}}, 
  year={2022},
  volume={9},
  number={18},
  pages={17323-17333},
  doi={10.1109/JIOT.2022.3156099}}

@article{martins2025delving,
  title={Delving Into Security and Privacy of Joint Communication and Sensing: A Survey},
  author={Martins and others},
  journal={IEEE Open J. Commun. Soc.},
  year={2025},
  publisher={IEEE}
}

@article{yildirim2025ofdm,
  title={{OFDM}-based {JCAS} under Attack: The Dual Threat of Spoofing and Jamming in {WLAN }Sensing},
  author={Yildirim and others},
  journal={IEEE Internet Things J.},
  year={2025},
  publisher={IEEE}
}

@article{ali2025cooperative,
  title={Cooperative {ISAC} Under Spoofing Attacks},
  author={Ali, Usman and others},
  journal={IEEE Wireless Communications Letters},
  year={2025},
  publisher={IEEE}
}

@INPROCEEDINGS{janjua-pilot-alloc-2022,
  author={Janjua, Muhammad Bilal and Memişoğlu, Ebubekir and Qaraqe, Khalid A. and Arslan, Hüseyin},
  booktitle={2022 IEEE Globecom Workshops (GC Wkshps)}, 
  title={Secure Pilot Allocation for Integrated Sensing and Communication}, 
  year={2022},
  volume={},
  number={},
  pages={1466-1471},
  doi={10.1109/GCWkshps56602.2022.10008617}}

@ARTICLE{xie2025-rcc-tag,
  author={Tan, Haijun and Xie, Ning and others},
  journal={IEEE J. Sel. Areas Commun}, 
  title={Physical Layer Authentication in Radar-Communication Coexistence Systems}, 
  year={2025},
  volume={},
  number={},
  pages={1-1},
  doi={10.1109/JSAC.2025.3612358}}

@article{tian2026-cfo-pla-isac,
title = {{CFO}-based physical-layer authentication for integrated sensing and communication under dynamic time resources},
journal = {Physical Communication},
volume = {74},
pages = {102977},
year = {2026},
issn = {1874-4907},
author = {Tuanwei Tian and others}
}

@ARTICLE{sankhe-rfimpair-2020,
  author={Sankhe, Kunal and Belgiovine and others},
  journal={IEEE Trans. Cogn. Commun. Netw.}, 
  title={No Radio Left Behind: Radio Fingerprinting Through Deep Learning of Physical-Layer Hardware Impairments}, 
  year={2020},
  volume={6},
  number={1},
  pages={165-178},
  doi={10.1109/TCCN.2019.2949308}}

@ARTICLE{liu-csi-pla-2018,
  author={Liu and others},
  journal={IEEE Trans. Mobile Comput.}, 
  title={Authenticating Users Through Fine-Grained Channel Information}, 
  year={2018},
  volume={17},
  number={2},
  pages={251-264},
  doi={10.1109/TMC.2017.2718540}}

@ARTICLE{aldaghri-chkey-2020,
  author={Aldaghri, Nasser and Mahdavifar, Hessam},
  journal={IEEE Trans. Inf. Forensics Security}, 
  title={Physical Layer Secret Key Generation in Static Environments}, 
  year={2020},
  volume={15},
  number={},
  pages={2692-2705},
  doi={10.1109/TIFS.2020.2974621}}

@article{xie2020survey,
  title={A survey of physical-layer authentication in wireless communications},
  author={Xie, Ning and Li, Zhuoyuan and Tan, Haijun},
  journal={IEEE Commun. Surveys Tuts.},
  volume={23},
  number={1},
  pages={282--310},
  year={2020},
  publisher={IEEE}
}

@article{tan2025asynchronous,
  title={Asynchronous Tag-based Physical-Layer Authentication in Wireless Communications},
  author={Tan and others},
  journal={IEEE Trans. Wireless Commun.},
  year={2025},
  publisher={IEEE}
}

@ARTICLE{yu2008-sup-pla,
  author={Yu, Paul L. and Baras, John S. and Sadler, Brian M.},
  journal={IEEE Trans. Inf. Forensics Security}, 
  title={Physical-Layer Authentication}, 
  year={2008},
  volume={3},
  number={1},
  pages={38-51},
  doi={10.1109/TIFS.2007.916273}
}

@ARTICLE{2024-chrysanidis-replay,
  author={Chrysanidis, Georgios and Liu, Yanwei and Argyriou, Antonios},
  journal={{IEEE} Access}, 
  title={A Replay Attack Against {ISAC} Based on {OFDM}}, 
  year={2024},
  volume={12},
  number={},
  pages={20998-21003},
  doi={10.1109/ACCESS.2024.3359680}}

@ARTICLE{du-xu-802.11bf-overview-2024,
  author={Du and others},
  journal={{IEEE Commun. Surveys Tuts.}}, 
  title={An Overview on {IEEE} {802.11bf}: {WLAN} Sensing}, 
  year={2025},
  volume={27},
  number={1},
  pages={184-217},
  doi={10.1109/COMST.2024.3408899}}

@ARTICLE{aldirmaz-6g-ISAC,
  author={Aldirmaz-Colak and others},
  journal={IEEE Access}, 
  title={A Comprehensive Review on {ISAC} for {6G}: Enabling Technologies, Security, and {AI}/{ML} Perspectives}, 
  year={2025},
  volume={13},
  number={},
  pages={97152-97193},
  doi={10.1109/ACCESS.2025.3573371}}

@standard{ieee80211be2024,
  author       = {{IEEE 802}},
  title        = {Part 11: Wireless {LAN} {MAC} and {PHY} Specifications—Amendment 7: Enhancements for Extremely High Throughput (EHT)},
  year         = {2024},
  number       = {IEEE Std 802.11be-2024},
  organization = {IEEE},
  month        = {September}
}
\end{document}